# StreamBED: Training Citizen Scientists to Make Qualitative Judgments Using Embodied Virtual Reality Training

**Alina Striner**
University of Maryland
College Park, MD 20852
Algo1001@umd.edu

**Jennifer Preece**
University of Maryland
College Park, MD 20852
preece@umd.edu

## Abstract
*Environmental citizen science frequently relies on experience-based assessment, however volunteers are not trained to make qualitative judgments. Embodied learning in virtual reality (VR) has been explored as a way to train behavior, but has not fully been considered as a way to train judgment. This preliminary research explores embodied learning in VR through the design, evaluation, and redesign of StreamBED, a water quality monitoring training environment that teaches volunteers to make qualitative assessments by exploring, assessing and comparing virtual watersheds.*

## ACM Classification Keywords
H.5.m Miscellaneous

## Introduction
Citizen science is a form of crowdsourcing that allows volunteers to collaborate with researchers on scientific data collection [1]. Environmental researchers spend years learning the nuances of collecting data, however volunteers often have limited experience in methodology. Volunteers that are thrown into data collection tasks without adequate training collect imprecise and biased data that negatively impacts science and advocacy work [1]. Poor training is further reflected in project retention; volunteers that are anxious or distrustful of their ability to perform research tasks tend not to participate over sustained periods [2].

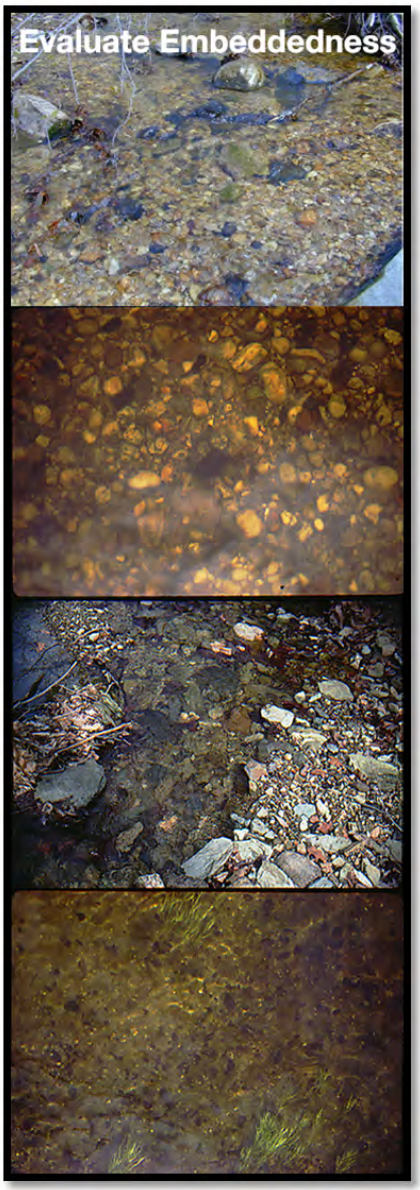

Figure 1: Background lecture material from the How to Read your Stream course[10]. This slide illustrates the difficulty of evaluating the embeddedness of cobbles in a stream. The EPA embeddedness scale is depicted in figure 3.

Training volunteers to make nuanced assessments requires a large demand on researcher time and resources. Current projects primarily train citizen science volunteers with passive training materials such as background reading modules, project instructions, and supporting materials [1]. Large-scale projects supplement reading materials with additional training workshops or class lectures [10], but most projects are under-resourced and understaffed [9], without means to conduct in depth training. Some effort has been made to educate citizen scientists, however training does not comprehensively address many of the challenges citizen scientists experience in the field or the needs of different age ranges and abilities.

**Qualitative Stream Assessment**
The focus of this research is on assessing stream water quality, a practice of sampling and analyzing stream conditions and water constituents [8]. This research draws on stream monitoring exposure and training with a Maryland water monitoring group [10] and consultations with an EPA water quality biologist [4].

*EPA Rapid Bioassessment Protocol*
The EPA Rapid Bioassessment Protocol (RBP) [8] is a standard monitoring protocol used by over 30 water quality groups across the US to visually assess habitat in order to monitor and report on the balance between development and environmental protection and to advocate for land use that protects regional watersheds. Although widely used, its subjectivity causes interpretation issues demonstrated in Table 1.

Some monitoring groups teach learners to make assessments through classroom PowerPoint lectures, however such passive training poorly addresses subjectivity issues (Table 1) and inadequately demonstrates differences within measures.

Table 1: Subjectivity issues in the RBP Bioassessment Protocol

| | |
|---|---|
| *How to interpret scales?* | The RBP protocol[8] suggests that measures should be assessed linearly, but experts suggest linearity varies between measures. For example, stream experts [4] interpret a stream with 25% or more embeddedness as poor because the environment is unsuitable for macro-invertebrate organisms even though the protocol considers the habitat suboptimal. |
| *How to evaluate within measure variation?* | Data collectors are asked to evaluate 100 meter stream cross sections, but how should users evaluate areas with significant variability? Experts [4] make holistic judgments of quality based using their experience, but new data collectors have no foundation with which to make judgments. |
| *How to evaluate related measures?* | Several measures of stream quality directly affect one another (e.g. stream bank stability affects sediment deposits). How should data collectors account for this in their assessments? |
| *How to interpret passage of time?* | 3 of 10 protocol measures ask users to evaluate transience of stream elements (*e.g.*, logs and cobble) and recency of human activity (*e.g.*, whether stream channel alteration occurred more or less than 20 years ago). How would users know how to judge the passage of time? |

For instance, a slide from the *How to Ready Your Stream* PowerPoint lecture [10] illustrates the challenge of evaluating particle embeddedness; EPA guidelines ask users to rate gravel, cobble and boulder embeddedness on a scale from 0 to 20 [8], but Figure 1 demonstrates that it is difficult to rate the 4 environments using the images alone.

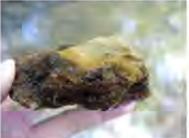

### StreamBED: Embodied Learning

With the growing popularity of virtual reality, studies have explored how VR may train objective judgments required in procedural tasks, such as surgery [7] and mining hazard response [5]. While VR has been found effective in procedural training, research has not substantively explored how the medium may be used for subjective holistic judgments.

In this paper, I present the initial design and evaluation of StreamBED, a water quality monitoring training integrated with VR that teaches volunteers to qualitatively assess virtual streams by maneuvering through virtual spaces and seeing streams from multiple angles and perspectives. The goal of this system is to give citizen scientists the interpretive skills that water biologists learn through years of monitoring experience.

Demonstrating that qualitative assessment skills can be trained in VR has significant implications for the water monitoring and citizen science community. Effective training has the potential to improve citizen science data quality, decrease cost, and improve recruitment and retention; in addition to creating meaningful training, tasks could be used to assess volunteer accuracy and precision, and to motivate citizen scientists to regularly participate in data collection [6].

### Water Quality Monitoring Training

Four measures from the RBP Protocol [8] were simulated in the Unity virtual environment: (1) *epifaunal substrate/available cover*, (2) *bank stability*, (3) *riparian vegetation zone width*, and (4) *channel alteration*. These measures were chosen based on their considerable impact on habitat (based on correlations to biological index scores)[3] and relative difficulty of qualitative assessment, as described in Table 1. The measures were modeled in diverse stream environments using the RBP protocol (see Figure 2).

Figure 2: Lecture slide from the *How to Read your Stream* course and Cobble embeddedness protocol scale.

***Platform*** The training environment was developed in the Unity 5 game engine and integrated with the Oculus Rift Head Mounted Display [HMD]. The environment was constructed using brushes, textures, assets and prefabs found in the Unity Asset store and online.

***Optimal and Poor Training*** consisted of participants navigating two tutorial environments: an optimal quality stream featuring an abundance of epifaunal substrate, a large riparian zone, no channel alteration and stable banks; and a poor quality stream with no epifaunal substrate, a lack of a riparian zone, high channel alteration and highly eroded unstable banks (see Figure 3). In addition to modeling these measures in the training, additional measures of stream quality were built in, including variability in pool depth, water clarity and vegetation diversity; this was designed to make the experience realistic and challenge participants to make assessments in context of other factors.

Participants first made topical observations of quality as they walked through the tutorials, then collected measure protocol and definition cards (Figure 5) and added them to a virtual field notebook (Figure 6). Picking up cards triggered a glowing ring to appear around a physical representation of each measure, guiding participants to study and interact with them.

***Assessment*** Having explored the optimal and poor tutorials, participants evaluated a virtual stream that exhibited a set of diverse quality characteristics; they picked up assessment cards (Figure 7) and used their experience in the tutorials and reference notebook to holistically assess the 4 measures. Since it is difficult to definitively assess a qualitative measure on a quantitative scale, participants received feedback based on the range that they chose on the scale.

### Method

Ten (10) participants were recruited from a pool of students taking classes at a large university who successfully passed a motion sickness pre-screening, and who consented to participate in the study. Participants interacted with the training using an Oculus Rift SDK2 and Xbox 360 Game Controller with standard key mappings.

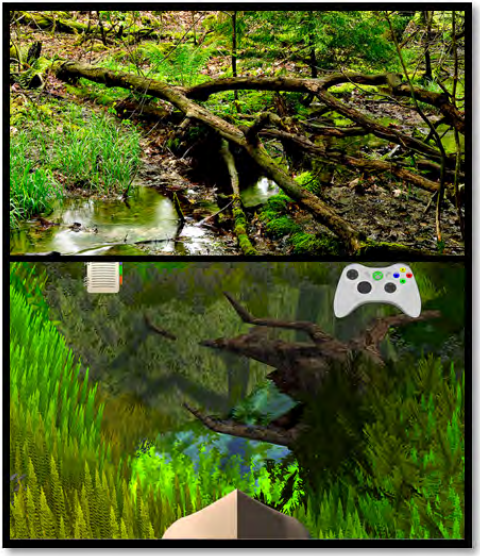

*Procedures*
The study consisted of participants exploring optimal and poor tutorial environments, making assessments of a realistic virtual stream, and making assessments of an actual stream on the university campus. During the study, participants assessed the usability of the system and their experience using a 7-point Likert scale.

**Virtual Training** After signing consent, participants filled out a background questionnaire and were introduced to the topics of citizen science and water quality monitoring. Participants were then guided through a short Xbox controller training allowing them to practice training interactions while being able to see the controller. After controller training, participants interacted with the tutorials and assessment using the Oculus Rift HMD and Xbox Controller; they were provided with water to drink and were encouraged to take breaks when they felt uncomfortable. Those who had trouble wearing the Oculus Rift (due to wearing glasses) visually explored the tutorials using the Oculus Rift, but completed tasks with a high definition monitor.

**Outdoor Data Collection** After training, participants walked outdoors to assess a stream on campus. During the walk to the stream, they evaluated their experience with the virtual training, and predicted their ability to accurately assess the measures they learned. At the stream, participants received a physical copy of the virtual reference notebook, and were asked to rate stream features based on the measures they learned, orally explaining their reasoning and decisions (outdoor data collection shown in Figure 4). After completing the outdoor assessment, participants completed a final questionnaire about the relevance of the virtual training to the physical task, and answered questions about their confidence and motivation to participate in future water-monitoring projects.

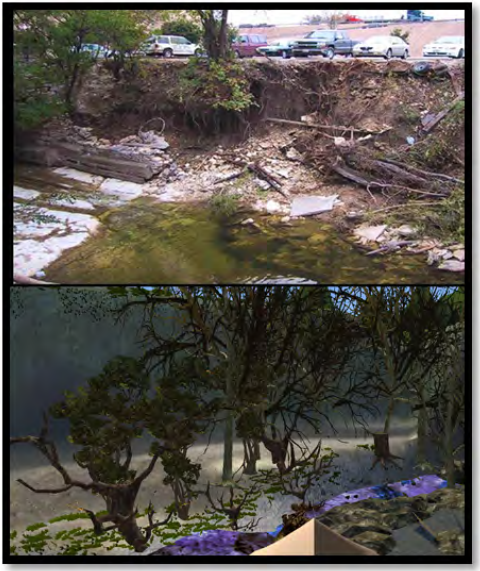

Figure 3: Comparable photos and Unity screen captures of optimal and poor environments

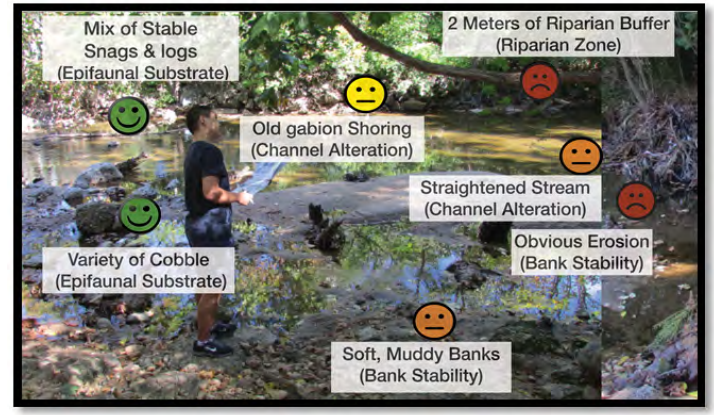

Figure 4: A participant making assessments of a stream displaying a synthesis of positive and negative characteristics.

*Analysis*
This research tested the preliminary design of StreamBED as a stream assessment-training tool. The small study population (n=10) limited the quantitative analysis to descriptive statistics, correlations, and scatterplot trends, so the limited evaluation strove to understand quantitative trends. Additionally, open coding was used to identify meaningful trends in user responses to training and data collection tasks; salient quotes were transcribed, grouped into overarching themes, and then organized into sub-themes. Identified themes were used to consider the effect of the training design on participant enjoyment, immersion, and motivation throughout the study.

**Quantitative Findings**
Study participants completed the virtual training in approximately 2 hours. Participants spent an average of 56 minutes on the optimal and poor environments, and 62 minutes assessing the virtual test stream. During the virtual assessment task, participants took approximately 14.47 tries to answer the 4 assessment questions (3.61 responses/question). Although participants had a large error rate, the assessment task was designed as a learning tool; during assessments, participants had to navigate different parts of the virtual stream to make accurate assessments, but it took time for them to associate the protocol with specific features of the environment. It is not surprising that participants had high error rates given the learning curve.

After training, participants made protocol assessments in an outdoor stream on the campus. Participant outdoor assessments were on average 2.37 points away from the correct response, a "gold standard" assessment made by the researcher and vetted by a water quality biologist[4]: participants were an average of 3.25 points away from the correct response on protocol scales ranging from 0 to 20, and were 1.93 points away from the gold standard on scales ranging from 0 to 10. Interestingly, there was a relatively strong positive correlation between total virtual training time, and participant's outdoor data collection scores ($r=.502$); participants who spent more on training also made assessments that were closer to the gold standard assessments. In contrast, there was a negative correlation between the amount of simulator sickness participants experienced, and the amount of time they spent on training ($r=-.338$). Participants who felt greater simulator sickness didn't spend as much time training, and also didn't collect as accurate data as participants who felt milder or no sickness.

There appears to also be a significant shift in average immersion from the beginning to the end of the study. At the beginning of the study, users predicted they would experience average immersion during training (4 on the Likert scale), however after training, immersion ratings shifted to 5.5, and after data collection, shifted all the way to 7.

**Open Coding Themes**
*Stories as Information*
Participants created stories that helped them explain virtual phenomena. One person joked, "*some crazed lumberjack came through here… And was…so buff that he picked up the logs with his hands and walked off with them,*" and another noted that "*human activity screwed the stream up so bad that it sort of grew back.*" Others saw the story of the changing landscape; "*There's not enough water here…some of the land that used to be the waterbed now [has] no water*." Another participant used the landscape to evaluate monitoring features; "*the rock around the channel tells me that this may be unnatural …everywhere else there is no rock.*"

*Virtual Surveying*
The virtual environment afforded participants with versatility that allowed them to meaningfully understand the environment. During an assessment task, one participant said that she was "*going to the higher ground to …have a better look,*" and several commented on their ability to be underwater, something hard to do in the real world. One participant noted that they could "*explore in a really intuitive way…I can walk real fast and look around in a more efficient way without getting muddy*." After training, a participant even commented that the VR training was more engaging than actual data collection; "*its similar,*" they said, "*but I cannot do as much as I could in the VR.*"

*Extraneous Information*
Participants were directed to make assessments using protocol measures, however they often took superfluous features under evaluation. Several participants commented that the water in the poor environment "*seems a little artificial*" and the grass in optimal environment "*was a little too green*." One participant asked about the variety of plants that were growing along the stream bank; "*[are these] supposed to be rice? Aquatic corn?*" Likewise, a participant commented than the environment looked poor because there weren't "*many leaves on [the trees].*"

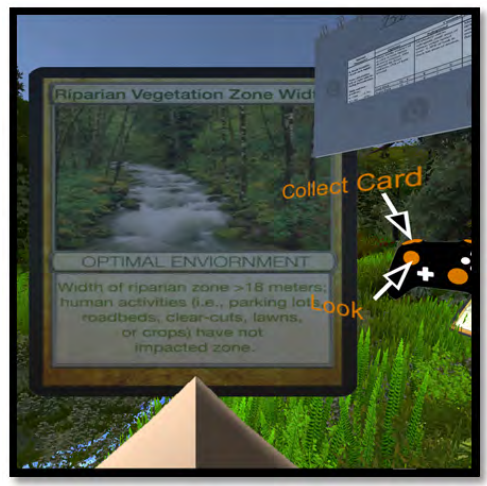

Figure 5: Sample card that participants collected as they explored tutorial environments. This card explains and demonstrates the protocol for an optimal Riparian Zone.

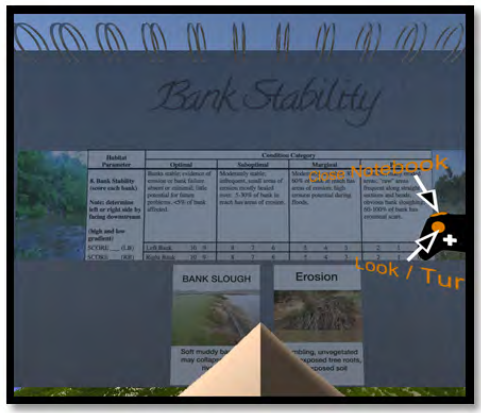

Figure 6: Filled out Bank Stability measure in the virtual notebook after participants collected protocol cards and definitions.

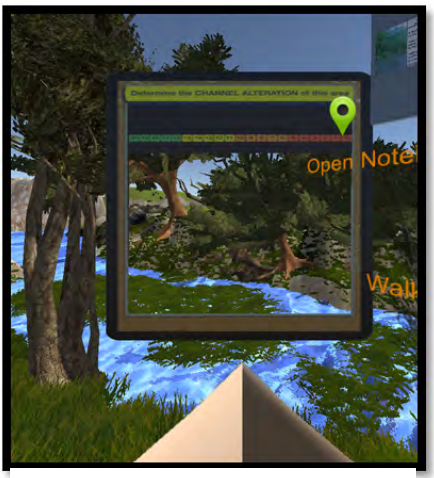

Figure 7: Assessment card asking participants to evaluate the channel alteration for the environment around them.

*Extreme Standards*
After seeing the optimal and poor tutorials, participants frequently judged moderate streams using extreme language. One participant described a test environment as "*closer to poor than optimal*" rather than as suboptimal or marginal, while another said one area is "*optimal…the left bank…but in the poor environment its something like…this side without grass*." Although the tutorials biased participants toward extremes, their answers suggested that they were using their holistic experiences to guide the assessments. For instance, a participant remarked, " *I won't [say] poor because there's not too much human being activities …it may be suboptimal… because it just don't look like…optimal*."

*Protocol Subjectivity*
Participants also have trouble with the subjectivity of the protocol. One participant didn't understand the differences within protocol subheadings. *"Why can't you choose 6 instead of 8 when they're both suboptimal?"* they asked, and another assumed that they were supposed to first choose a category (e.g., marginal) before choosing a numeric answer. One participant even wondered if the scores were percentages. *"6 means 60% and 7 means 70%?"* Even when participants understood the protocol, they commented on its subjectivity. The protocols "*had weird mappings…[that] felt unnatural…why can't you do 4 levels, like optimal, suboptimal, marginal, poor?*"

## Conclusions and Implications
These early findings unveil an opportunity to meaningfully train citizen scientists to make holistic water quality assessments through embodied learning, physical exploration of the virtual environment. Although the main themes suggest that virtual training has positive implications, the study revealed several challenges that need to be addressed to effectively train participants; Study themes revealed that participants used their experiences to make assessments, but also needed concrete guidance to identify key features of the environment. Additionally, participants needed clear explanations of how to use the protocol scale, and substantial feedback on their responses.

*Future Work*
This study was an initial design and evaluation of virtual water quality training to teach holistic assessment with embodied learning. In addition to iterating on the current training design, multisensory interactions should explored in parallel to VR. Could sounds or smells provide virtual training cues? As well as investigating the role of multisensory design, future research should also consider the impact of controller interactions, story, and collaborative learning on training effectiveness.

This study's focus was college-age participants, however citizen science has historically drawn older, more affluent volunteers[1]. Future work should thus also consider how this participant pool reacts to virtual training. As well as addressing the needs of different populations, future work might additionally address whether virtual training is appropriate for a broad scope of citizen science research, such as conservation or education projects.

## Acknowledgements
Thanks to Dr. Gregory Pond from the US EPA, and to Ms. Cathy Wiss from the Audubon Society for their generous time and feedback.